# Machine Learning in Finance-Emerging Trends and Challenges

Jaydip Sen*[1], Rajdeep Sen[2], and Abhishek Dutta[3]
[1]Department of Data Science, Praxis Business School, Kolkata, INDIA
[2]Independent Researcher and Financial Analyst
[3]School of Computing and Analytics, NSHM Knowledge Campus, Kolkata, India.
*Corresponding author email: jaydip.sen@acm.org.

## 1. Introduction

The paradigm of machine learning and artificial intelligence has pervaded our everyday life in such a way that it is no longer an area for esoteric academics and scientists putting their effort to solve a challenging research problem. The evolution is quite natural rather than accidental. With the exponential growth in processing speed and with the emergence of smarter algorithms for solving complex and challenging problems, organizations have found it possible to harness a humongous volume of data in realizing solutions that have far-reaching business values.

Financial services, banking, and insurance remain one of the most significant sectors that has a very high potential in reaping the benefits of machine learning and artificial intelligence with the availability of rich data, innovative algorithms, and novel methods in its various applications. While the organizations have only skimmed the surface of the rapidly evolving areas such as deep neural networks and reinforcement learning, the possibility of applying these techniques in many applications vastly remains unexplored. Organizations are leveraging the benefits of innovative applications of machine learning in applications like customer segmentation for target marketing of their newly launched products, designing optimal portfolio strategies, detection, and prevention of money laundering and other illegal activities in the financial markets, smarter and effective risk management is credit, adherence to the regulatory frameworks in finance, accounts, and other operations, and so on. However, the full capability of machine learning and artificial intelligence still remains unexplored and unexploited. Leveraging such capabilities will be critical for organizations to achieve and maintain a long-term competitive edge.

While one of the major reasons for the slow adoption of AI/ML models and methods in financial applications is that the algorithms are not well known and there is an inevitable trust deficit in deploying them in critical and privacy-sensitive applications, the so-called "black-box" nature of such models and frameworks that makes analysis of their internal operations in producing outputs and their validations also impede faster acceptance and deployment of such models in real-world applications.

This introductory chapter highlights some of the challenges and barriers that organizations in the financial services sector at the present encounter in adopting machine learning and artificial intelligence-based models and applications in their day-to-day operations.

The rest of the chapter is organized as follows. Section 2 presents some emerging applications of machine learning in the financial domain. Section 3 highlights emerging computing paradigms in finance. Some important modeling paradigms in the era of machine learning and artificial intelligence are discussed in Section 4. Section 5 discusses some new challenges and barriers currently faced by the financial modelers. Some emerging risks, new choices, and modern practices in financial modeling are presented in Section 6. Finally, Section 7 concludes the chapter.

## 2. Emerging application of machine learning in finance

With the increasing availability and declining cost for complex models executing on high-power computing devices exploiting the unlimited capacity of data storage, the financial industry is geared up to exploit the benefits of machine learning to leverage a competitive business edge. While some of the use cases have already found their applications in the real world, others will need to overcome some existing business and operational challenges before they are deployed. Some of the applications are mentioned below.

**Risk modeling:** One of the major applications of AI/ML models and algorithms is in the extensive domain of risk modeling and management [1]. While on one hand, the risk modeling credit and market is a critical application of machine learning, on the other hand, a non-finance application such as operational risk management, compliance, and fraud management is also quite important. The majority of the classification approaches and modeling techniques in machine learning such as binary logistic regression, multinomial logistic regression, linear and quadratic discriminant analysis, and decision trees, etc., are the foundational building blocks of applied modeling in the real world. However, in data science applications, the availability of data and its richness play a pivotal role. Hence, in data-rich applications such as credit risk modeling and scoring, designing mortgage schemes, the AI/ML models have already made substantial inroads in comparison to scenarios such as low default credit portfolios for well-known parties that lack the availability of data. Fraud analytics remains another intensive area of AI/ML applications in the non-financial domain.

**Portfolio management:** The portfolios are designed based on the recommendations of smart algorithms that optimize various parameters with return and risk being the two most important ones [2]. Using the information provided by the users such as their ages of retirement, amount of investment, etc., and other associate details such as their current ages, current assets at hand, the algorithm allocate the invested amount into diverse asset classes in order to optimize the return and the risk associated with the portfolio. Once an initial allocation is made, the algorithm continuously monitors the market environment and changes the allocation so that the portfolio remains at its optimized level always. These AI-enabled portfolio managers, known as the Robo-advisors are increasingly being used in real-world portfolio design due to their superior adaptability and optimization skill to their human counterparts.

**Algorithmic trading:** Algorithmic trading exploits the use of algorithms to carry out stock trading in an autonomous manner with the minimal human intervention [3]. Invented in 1970, algorithmic trading deploys automated pre-programmed stock trading instructions that can be executed very fast and at a very short time interval (i.e., at a very high frequency) to optimize trading returns and objectives. Machine learning and artificial intelligence have pushed algorithmic trading into a new dimension where not only advanced trading strategies can be made very fast but also deep insights can be made into the stock price and overall market movements. While most hedge funds and financial organizations do not make their trading strategies public, it is well known that machine learning, of late, is playing an increasingly important role in calibrating high-frequency, high-volume trading decisions in real-time for critical applications.

**Fraud detection and analysis:** Fraud detection and analysis is one of the most critical machine learning applications in the finance industry [4]. There is an increased level of security and privacy risk associated with sensitive information both from the organization and personal front due to ubiquitous availability of connectivity, high computational power in devices, an increased amount of data stored and communicated online. These issues have changed the way online fraud analysis and detection are being made. While detection in earlier days used to depend on matching a large set of complex rules, the recent approaches are largely based on the execution of learning algorithms that adapts to new security threats making the detection process more robust while being agile.

**Loan and insurance underwritings:** Loans, credit, and insurance underwriting is also an area where the large-scale deployment of machine learning models can be made by financial institutions for achieving competitive advantage [5]. At large banks and insurance firms, the availability of historical data of consumers, financial lending/borrowing information, insurance outcomes, and default-related information in paying debts, can be leveraged in training robust machine learning models. The learned patterns and trends can be exploited by the learning algorithms for lending and underwriting risks in the future to minimize future defaults. The deployment of such models can lead to a paradigm shift in business efficiency and profit. However, at present, there is a limited utilization of such models in the industry as their deployments are largely confined within large financial institutions.

**Financial chatbots:** Automation in the finance industry is also an outcome of the deployment of machine learning and artificial intelligence. Accessing the relevant data, machine learning models can yield an insightful analysis of the underlying patterns inside them that helps in making effective decisions in the future. In many cases, these models may provide recommended actions for the future so that the business decision can be made in the most efficient and optimum way. AI-based systems in financial applications also have the ability to minimize their errors learning fast from their past actions that also reduce wastages of precious resources including time. AI chatbots provide an effective way of interaction with the customers while automating many routine tasks in a financial institution [6].

**Risk management:** Machine learning techniques are revolutionizing the way corporates handle the risks associated with their operations. Risk management examples are diverse and infinite ranging from deciding about the amount a bank should lend a customer or how to improve compliance to a process or the way risk associated with a model can be minimized [7].

**Asset price prediction:** Predicting future asset prices is one of the most challenging tasks that finance professionals have to do frequently. The price of an asset is affected by numerous factors driven by the market including speculative activities. In the classical approach, an asset price is determined by analyzing historical financial reports and past market performances. With rich data available, of late, machine learning-based models have started playing significant roles in predicting future asset prices in a robust and precise manner [8].

**Derivative pricing:** In the traditional approach, derivative pricing models are designed on numerous assumptions which do not hold good in the real world. These assumptions attempt to establish an empirical relationship among several predictor variables such as the strike price, maturity time, and option type, and the target variable which is the price of the derivative. Machine learning models have got rid of the requirement of such assumptions as they attempt to fit in the best possible function between the predictors and the target by minimizing the error. Accuracy and the minimal time needed for the deployment of the models in real-world use cases make machine learning the most impactful paradigm in the task of derivative pricing [9].

**Money laundering:** As per a report published by the United Nations, it is estimated that 2% to 5% of the world's aggregated GDP is accounted for the amount of money being laundered worldwide. Machine learning models can find applications in detecting money laundering activities with minimal false-positive cases [10].

## 3. Emerging paradigms in computing in finance

Several new capabilities and approaches and frameworks in machine learning, data science, and artificial intelligence have become available to the modelers and engineers for all disciplines including finance professionals and researchers. Some of them are as follows.

**Virtual agents:** The machine learning paradigm will witness the increasing deployment of agents in various tasks. These agents have the capability of performing complex data mining tasks through a large set of policy rules, defined procedures and regulations, and provide automated responses to queries.

**Cognitive robotics:** The robots in the cognitive domain have the power of automating several tasks which are currently done by humans. This automation comes with an additional level of sophistication, speed, and precision in performing the tasks.

**Text analytics:** The applications of sophisticated algorithms, frameworks, and models of natural language processing in analyzing voluminous and complex financial contracts and documents help processing and decision making faster and more accurately with minimal associated risks.

**Video analytics:** Advancements in the fields of computer vision, image processing, speech processing, and speech recognition together with the exponential growth in hardware capabilities have led to very promising progress in compliance, audit, model validation in various financial applications including automated generation and presentation of financial reports.

## 4. Emerging trends in modeling techniques

With the increasing proliferation of machine learning models in innovative applications in the financial industry, some computing and modeling paradigms will find more adoption. Some of them are as follows.

**Sparsity-aware learning:** Sparsity-aware learning has evolved as an alternative model regularization approach to address a number of problems that are usually encountered in machine learning [11]. Considerable effort has been spent in designing schemes such frameworks in an iterative manner in solving estimation tasks of model parameters avoiding overfit. Sparsity-aware learning systems are well-suited in financial modeling applications leading to extremely robust and accurate models for various applications in finance.

**Reproducing Kernel Hilbert Spaces:** Reproducing Kernel Hilbert Spaces (RKHS) is essentially a Hilbert space function that evaluates a continuous function in the linear space [12]. These functions find important applications in statistical learning as every functional representation in RKHS represents minimization of an empirical function embodying the associated risk, and the representation is made as a linear combination of the data points in the training set transformed by the kernel function. Accordingly, RKHS has a very high potential in risk modeling and evaluation in finance.

**Monte Carlo simulation:** This method of modeling provides the modeler with a large range of possible outcomes and probabilities that they will occur for any choice of action that is taken. It is used in a diverse set of domains like finance, energy, project management, and monitoring, research and development, and insurance. It performs risk analysis by designing models of possible results by substituting a range of values – a probability distribution – for any factor that has inherent uncertainty. The ability in handling uncertainty makes this approach particularly popular in modern-day modeling in finance [13].

**Graph theory:** Multivariate financial data pose a very complex challenge in processing and visualization in addition to being difficult in modeling. Graph theory provides the modeler with a very elegant, efficient, and easily interpretable method of handling multivariate financial data [14].

**Particle filtering:** It is a method of modeling nonlinear and non-Gaussian systems with a very high level of precision. Its ability to handle multi-modal data makes it one of the most effective and popular modeling techniques in many fields including finance [15]. Stated in simple words, particle filtering is a technique for identifying the distribution of a population that has a minimum variance by identifying a set of random samples traversing through all the states in order to obtain a probability density function that best fits into the original distribution and then substituting the integral operation on the function by the mean of the sample.

**Parameter learning and convex paths:** While optimization methods have been proved to be very effective in training large-scale deep neural networks involving millions of parameters, the regularization of these methods has become of paramount importance for proper training of such networks [16]. Accordingly, intensive work has been also carried out in estimating the biases associated with the optimum value of the objective function arrived at by the algorithms. The estimation of such biases provides the modeler with an idea about the degree of inaccuracy in the models for critical applications including financial modeling.

**Deep learning and reinforcement learning:** The application of machine learning in finance has largely been manifested in the form of models built on deep neural network architecture and smarter algorithms for the optimization and training of such networks. Reinforcement learning-based models have made the automation of such models a reality. A vast gamut of applications, such as algo trading, capital asset pricing, stock price prediction, portfolio management can be very effectively designed and executed using deep learning and reinforcement learning frameworks [17-26].

## 5. New challenges in financial modeling

Despite the numerous opportunities and advantages that machine learning and artificial intelligence-based applications are likely to provide to the financial sector, there will be some initial challenges and barriers too. This section highlights some of the challenges as follows.

**Data challenges:** While the availability of data in finance is quite plenty, the time series data in finance (e.g., stock prices) are quite small in size for data-hungry machine learning and deep learning models. Models built on limited time series data are naturally less trained and improperly designed. The result is a sub-optimal performance of the models. Another problem in finance is that financial data cannot be synthesized unlike images in the fields of computer vision and image processing. Since finance data cannot be synthesized, one has to wait for financial data to be produced in the real world before using them in model training and validation. The third challenge with financial data is the high level of noise associated with high-frequency trading data. Since high-frequency data in finance are invariably associated with a high level of noise, the machine learning models trained on such noisy data are intrinsically imprecise. Data evolution in finance poses the fourth challenge. Unlike data in most other fields, where the data features do not change with time, the features in financial data in harmony with financial markets evolve and change with time. This implies that financial variables will not carry the same meaning and significance over a long period, say one decade. The changes in the semantic meaning and significance of financial variables make it particularly difficult for the machine learning model to derive a consistent explanation and learning over a reasonably long period of time [27].

**Black-box nature of the models:** Machine learning and artificial intelligence-based models are black-box in nature [28]. In these models, while the outputs from the model are available and most of the time, they are easily interpretable, the biggest shortcoming is their lack of power of explanation of the output. In many critical applications in finance, mere outputs are not sufficient, and strong logical support for explaining the output is mandatory to instill sufficient confidence in the minds of the decision-makers. In absence of such explainable characteristics of the machine learning models, it will always remain a difficult job for the modelers to advocate the suitability of such models in critical business use cases.

**Validation challenges of the models:** Due to their higher complexity and opaqueness in operation, the machine learning models pose some significant challenges to risk management and validation [29]. While the regulators demand the machine learning models to comply with the SR 11-7 and OCC 2011-12 standards of risk management, the optimum execution of the models may not be possible if all those guidelines are to be strictly adhered to. Model risk is an event that occurs when a model is designed following its intended objective but it introduces errors while in execution yielding inaccurate results from the perspective of its design and business use case. Another manifestation of model risk can happen when a model is built and deployed inaccurately or with faults without proper knowledge about its limitations and shortcoming.

**Challenges in model testing and outcome analysis:** The performance of a model and its accuracy in testing are evaluated by outcome analysis [29]. Since the neural network model has a natural tendency to overfit or underfit the data based on the training, it is imperative on the part of the model evaluation team to address the bias-variance trade-offs in the training and validation process. Since the traditional k cross-validation procedure used in backtesting of predictive models does not work effectively for machine learning model validation, the machine learning model validators should take particular care in carrying out normalization and feature selection before model training and validation. Validation loss and accuracy should be strictly monitored and analyzed to ensure that no overfitting or underfitting is present in the model before it is sent to the production phase. Neural network models are also difficult to evaluate on their sensitivity analysis as these models lack explainability. Since establishing a functional relationship between the explanatory variables and the target variable in a neural network model is a difficult task unlike statistical models, sensitivity analysis between the inputs and the output may involve a computationally involved challenge while the results of the analysis may be quite complex.

**Challenges with models designed by vendors:** As per the requirements specified in SR 11-7 and OCC 2011-12 standards, models supplied by vendors are required to adhere to the same rigor as internally developed models [29]. However, in many practical situations, due to proprietary restrictions testing of vendors supplied models becomes a challenge. For vendor-supplied models, banks and financial institutions will have to mostly rely on softer forms of validation. The softer form of validation may include periodic review of model performance and conceptual soundness, stringent assessment of model customization, review of the development process, and applicability of the model in the portfolio of operations of the bank.

## 6. Emerging risks, new choices, and modern practices

In all domains including finance, the major cause that contributes to the risk in a machine learning model is the complexity associated with the model. The machine learning algorithms are intrinsically very complex as they work on voluminous and possibly unstructured data such as texts, images, and speech. As a consequence, training of such algorithms demands sophisticated computing infrastructure and a high level of knowledge and skill on the part of the modelers. However, countering such complexities with overly sophisticated models can be very counterproductive. Instead of adopting a complex approach, the banks and financial institutions should understand the risks associated with their business and operations and manage them with simple model-validation approaches. The issues that will dominate the risk management landscape for machine learning models in the financial industry are – (i) interpretability of the models, (ii) model bias, (iii) extensive feature engineering, (iv) importance of model hyperparameters, (v) production readiness of models, (vi) dynamic calibration of models, and (vii) explainable artificial intelligence. These issues are briefly discussed below.

**Model interpretability:** Since all machine learning models are essentially black boxes, their interpretability requirements have to be determined by the banks and financial institutions based on the risk associated with their use cases [28]. While some use cases may demand a highly granular and transparent interpretation of the model's working, the requirements for some use cases may not be so stringent. For example, in credit granting by banks to its customers, there may be a clear explanation required to be given by a machine learning model in cases where such credits are refused by the banks. However, another use case that involves minimal risk to the bank's operations such as recommendation of a product sent to the mobile app installed on a customer's hand-held device may not demand any understanding of the reason why the model has made such recommendation.

The model validation process should make sure that models comply with their intended objectives and policies. Despite being black-box in nature, machine learning models come with various provisions for their interpretation. Depending on the type of the model, these approaches may be widely different as discussed below.

Models like linear regression which are not only linear but monotonic in their behavior, the coefficients associated with the explanatory variables exhibit their respective influence on the target variable in the model.

Ensemble models such as ada boosting or gradient boosting, are nonlinear but monotonic in their behavior. In these monotonic models, if the explanatory variables are restricted in their values, the restriction will cause either an increase or a decrease in the value of the target variable. This monotone nature of the models simplifies the contributions of the explanatory variables in the predicted values of the target variable by the model.

The complex deep neural network-based models which are not only nonlinear but also non-monotonic have methods like *Shapley additive explanations* (SHAP), and *local interpretable model agnostic explanations* (LIME) for their global and local interpretability.

**Model bias:** Machine learning models are susceptible to four different types of bias. These biases are, (i) sample bias, (ii) bias in measurement, (iii) bias due to algorithms, and (iv) bias towards or against a particular category of people [29]. For example, the algorithm used in building a random forest model has a bias towards input features that have a more distinct set of values. For example, a model built on a random forest algorithm for assessing potential money laundering activities is likely to be biased towards the features with a higher number of levels in its categories (e.g., occupation), while features having a lower number of categorical levels (e.g., country) might be better suited to detect money laundering activities. To counter the issues pertaining to algorithmic bias, the validation processes must be equipped with the ability to select the most appropriate algorithm in a given context. Feature engineering and designing challenger models are some of the methods to counter algorithmic biases in models. The bias against or in favor of a particular group of people can be avoided by defining a fair set of policies by the banks and financial institutions. Models should be tested for fairness for various cases, and necessary corrections should be made in case aberrant outputs are produced by the models.

**Extensive feature engineering:** The task of feature engineering involves much more complications in machine learning and deep learning models than classical statistical models. The factors that contribute to the increased complexity in feature engineering in machine learning are as follows. The first and the most obvious reason is the large number of features involved in machine learning models. First, machine-learning models usually involve a very large number of input variables. The second reason is due to the increased use of unstructured data input in machine learning models. Unstructured data like text, images, and speech involve an enormously large number of features after data preprocessing which add to the complexity in modeling. The use of commercial-off-the-shelf frameworks in modeling such as AutoML has also led to an increased number of features as these frameworks automatically generate derived features from the raw data for providing a better fit of the models into the data in the training phase. While such an approach leads to better training, it is very likely to yield overfitted models in most practical use cases. It is imperative for the banks to have a robust feature engineering strategy in place for mitigating their operational and business risks. The feature strategy is likely to be different for a diverse set of applications. In the case of highly critical and risk-intensive applications like the evaluation of credit-worthiness of customers, every single feature in the model needs to be assessed very closely. On the other hand, for routine applications involving low risks, an overall review of the feature engineering process involving a robust data wrangling step may be sufficient.

**Model hyperparameters:** In machine learning models, the algorithms have parameters associated with them that are not parameters of the models. For example, the number of levels in the constituent decision trees of a random forest (also known as the depth of the trees), or the number of hidden layers in the architecture of a deep learning model must be decided before the models can be trained. Stated in a different way, the parameters are not determined from the training data of the models, and these are called hyperparameters of the models. The values of hyperparameters in machine learning models are most often determined by trial-and-error methods or brute force search methods like grid search. The absence of an exact and efficient way of finding the optimum values of the hyperparameters makes designing machine learning and deep learning models a particularly challenging task as a wrong choice of values of the hyperparameters can lead to imprecise models. Of late, banks and other financial institutions are depending more on sophisticated binary classification models built on support vector machines with the additional capability of text analytics for analyzing customer complaints. However, these models will be difficult to generalize for a multitude of applications as they will be sensitive to the kernel used in the training.

**Production readiness of models:** Unlike statistical models which are designed as an aggregation of rules of codes to be executed in a production system, machine learning models are built on algorithms requiring intensive computations most of the time. However, in many situations, financial model developers ignore this important point and tend to build overly complex models only to find later that the production systems of the banks are unable to support such complexity. A very simple example could be a highly complex deep learning model built for detecting frauds in financial transactions that is unable to satisfy the latency constraint in its response. To ensure that the machine learning models are production-ready, the validation step should make a reliable estimation of the volume of data that the model will require to process on the production architecture [30].

**Dynamic calibration of models:** There is an adaptable class of machine learning models that has the ability to dynamically modify their parameters based on the patterns in the data [31]. An obvious example is a model built on the reinforcement learning approach. While the autonomous learning of such models is a great boon, there is also a new type of risk associated with such models. In absence of sufficient external control and with too much emphasis on learning from the short-term patterns on the data, the long-term performance of the models may be adversely affected. The consequence is an additional complexity in financial modeling – when to recalibrate the model dynamically? The dynamic recalibration time will also need to be adapted for different applications like algorithmic trading, creditworthiness determination, etc. A comprehensive validation process should now include a set of policies that will guide the modelers in evaluating dynamic recalibration so that the model can adhere to its intended objectives while appropriate controls are in place for ensuring that risks are mitigated when they emerge. This is surely not going to be an easy task as it will involve the complex job of thresholding, identifying the health of the models based on their critical metrics of performance on out-of-sample data.

**Explainable artificial intelligence:** In the explainable artificial intelligence paradigm, an AI program scans through the code of another AI program and attempts to explain the operating steps and the output yielded by the latter program. This approach can be exploited to adjust the values of the explanatory variables in a predictive model so that the desired value of the target variable (i.e., the output of the model) is obtained [32]. In this way, explainable AI provides a very easy and elegant way to record, analyze and interpret the learning method of a complex model and for repeating the same in the future. Although the computing paradigm is still in research labs, its adoption in a commercial environment especially in the financial industries is not far away.

## 7. Conclusion

In the days to come, the financial industry will show increasingly more reliance on machine learning and artificial intelligence-based emerging methods and models to leverage competitive advantages. While the regulatory and compliance will evolve into a more standardized framework, machine learning will continue to provide the banks and other financial institutions more opportunities to explore and exploit emerging applications, while being more efficient in delivering the existing services. While the emerging techniques discussed in the chapter will play their critical roles in mitigating future risks in models, they will also guide the authorities in designing effective regulations and compliance frameworks in risk-intensive applications like creditworthiness assessment, trade surveillance, and capital asset pricing. The model validation process will increasingly be adapted to mitigate machine learning risks, while considerable effort and time will be spent in fine-tuning the model hypermeters in handling emerging applications. However, banks will have more opportunities to deploy the models in a large gamut of applications, gaining competitive business advantages and mitigating risks in operations.

## References

1. Paltrinieri, N., Comfort, L., Reniers, G. Learning about risk: Machine learning for risk assessment. *Safety Science*, **118**(2019), 475-486, 2019. DOI: 10.1016/j.ssci.2019.06.001.
2. Sen, J., Mehtab, S. A comparative study of optimum risk portfolio and eigen portfolio on the Indian stock market. *Int. J. of Business Forecasting and Mktg. Intl.*, Inderscience, Paper ID: IJBFMI-90288, 2021. (Accepted for publication)
3. Lei, Y., Peng, Q., Shen, Y. Deep learning for algorithmic trading: Enhancing MACD strategy. In: *Proc. of the 6th Int. Conf. on Comptg. and Artificial Intelligence*, pp. 51-57, April 2020, Tianjin, China. DOI: 10.1145/3404555.3404604.
4. Dornadula, V. N., Geetha, S. Credit card fraud detection using machine learning algorithms. *Procedia Computer Science*, **165**(2019), 631-641, 2019. DOI: 10.1016/j.procs.2020.01.057.
5. Eling, M., Nuessl, D., Staubli, J. The impact of artificial intelligence along the insurance value chain and on the insurability of risks. *Geneva Paper on Risk and Insurance-Issues and Practices*, 2021. DOI: 10.1057/s41288-020-00201-7.
6. Yu, S., Chen, Y., Zaidi, H. AVA: A financial service chatbot based on deep bidirectional transformers. *Frontiers in Applied Mathematics and Statistics*, **7**:604842, 2021. DOI: 10.3389/fams.2021.604842.
7. Leo, M., Sharma, S., Maddulety, K. Machine learning in banking risk management: A literature review. *Risks*, **7**(1), 2019. DOI: 10.3390/risks7010029.
8. Gu, S., Kelly, B., Xiu, D. Empirical asset pricing via machine learning. *The Review of Financial Studies*, **33**(5), 2233-2273, 2020. DOI: 10.1093/rfs/hhaa009.
9. Ye, T., Zhang, L. Derivatives pricing via machine learning. *Journal of Mathematical Finance*, **9**(3), 561-589, 2019. DOI: 10.4236/jmf.2019.93029.
10. Zand, A., Orwell, J., Pfluegel, E. A secure framework for anti-money-laundering using machine learning and secret sharing. In: *Proc. of Int. Conf. on Cyber Sec. and Protection of Digital Services*, pp. 1-7, Dublin, Ireland, June 15-19, 2020. DOI: 10.1109/CyberSecurity49315.2020.9138889.
11. Theodoridis, S., Kopsinis, Y., Slavakis, K. Sparsity-aware adaptive learning: A set theoretic estimation approach. *IFAC Proceedings Volumes*, **46**(11), 748-756, 2013. DOI: 10.3182/20130703-3-FR-4038.00157.
12. Theodoridis, S. Learning in Reproducing Kernel Hilbert Space. Chapter 11, Ed. Theodoridis, S. *Machine Learning*, pp. 509-583, Academic Press, 2015. DOI: 10.1016/B978-0-12-801522-3.00011-2.
13. Zhang, Y. The value of Monte Carlo model-based variance reduction technology in the pricing of financial derivatives. *PLoS ONE*, **15**(2): e0229737, 2020. DOI: 10.1371/journal.pone.0229737.

14. Cardoso, J. V. D. M, Palomar, D. P. Learning undirected graphs in financial markets. In: *Proc. of 54th Asilomar Conf. on Signals, Systems, and Computers*, pp. 741-745, Pacific Grove, CA, USA, 2020. DOI: 10.1109/IEEECONF51394.2020.9443573.
15. Fukui, T., Sato, S., Takahashi, A. Style analysis with particle filtering and generalized simulated annealing. *International Journal of Financial Engineering*, **4**(2-3), 1750037, 2017. DOI: 10.1142/S2424786317500372.
16. Boyd, S., Vandenberghe, L. *Convex Optimization*. Cambridge University Press, Cambridge, UK, 2013. Online ISBN: 9780511804441. DOI: 10.1017/CBO9780511804441.
17. Sen, J., Mehtab, S. A time series analysis-based stock price prediction using machine learning and deep learning models. *Int. J. of Business Forecasting and Mktg, Intelligence*, 6(4), 272-335, 2020. DOI: 10.1504/IJBFMI.2020.115691.
18. Sen, J., Mehtab, S. Accurate stock price forecasting using robust and optimized deep learning models. In: *Proc. of IEEE Int. Conf. on Intelligent Tech (CONIT)*, June 25-27, 2021, Hubballi, India. DOI: 10.1109/CONIT51480.2021.9498565.
19. Sen, J., Dutta, A., Mehtab, S. Profitability analysis in stock investment using an LSTM-based deep learning model. In: *Proc. of 2nd IEEE Int. Conf. for Emerging Technologies (INCET)*, pp. 1-9, May 21-23, 2021, Belagavi, India. DOI: 10.1109/INCET51464.2021.9456385.
20. Mehtab, S., Sen, J. Analysis and forecasting of financial time series using CNN and LSTM-based deep learning models. In: *Proc. of 2nd IEEE Int. Conf. on Advances in Distributed Comp. Mach. Learning (ICADCML)*, Jan 15-16, Bhubaneswar, India. (In press)
21. Mehtab, S., Sen, J., Dasgupta, S. Robust analysis of stock price time series using CNN and LSTM-based deep learning models. In: *Proc. of 4th IEEE Int. Conf. on Electronics, Communication and Aerospace Tech (ICECA)*, pp. 1481-1486, Nov 5-7, 2020, Coimbatore, India. DOI: 10.1109/ICECA49313.2020.9297652.
22. Mehtab, S., Sen, J. Stock price prediction using CNN and LSTM-based deep learning models. In: *Proc. of IEEE Int. Conf. on Decision Aid Sciences and Applications (DASA)*, pp. 447-453, Nov 8-9, 2020, Bahrain. DOI: 10.1109/DASA51403.2020.9317207.
23. Mehtab, S., Sen, J., Dutta, A. Stock price prediction using machine learning and LSTM-based deep learning models. In: *Machine Learning and Metaheuristics Algorithms, and Applications (SoMMA)*, 2020, pp. 88-106, Springer Nature, Singapore. DOI: 10.1007/978-981-16-0419-5_8.
24. Mehtab, S., Sen, J. Stock price prediction using convolutional neural networks on a multivariate time series. In: *Proc. of 3rd Nat. Conf. on Machine Learning and Artificial Intelligence (NCMLAI)*, New Delhi, India, Feb 1, 2020. DOI: 10.36227/techrxiv.15088734.v1.
25. Mehtab, S., Sen, J. A robust predictive model for stock price prediction using deep learning and natural language processing. In: *Proc. of 7th Int. Conf. on Business Analytics and Intelligence (BAICONF)*, Dec 5-7, 2019, Bangalore, India. DOI: 10.36227/techrxiv.15023361.v1.
26. Sen, J. Stock price prediction using machine learning and deep learning frameworks. In: *Proc. of 6th Int. Conf. on Business Analytics and Intelligence (ICBAI)*, Dec 20-21, 2018, Bangalore, India.
27. L. Cao. AI in finance: Challenges, techniques, and opportunities. 2021. Available online at: https://arxiv.org/abs/2107.09051 (accessed on Oct 7, 2021)
28. Rudin, C. Stop explaining black-box machine learning models for high stakes decisions and use interpretable models instead. *Nature Machine Intelligence*, **1**, 206-215, 2019. DOI: 10.1038/s42256-019-0048-x.
29. Huang, J., Chai, J. Deep learning in finance and banking: A literature review and classification. *Frontiers of Business Research in China*, **14**, Art Id: 13, 2020. DOI: 10.1186/s11782-020-00082-6.
30. Studer, S., Bui, T. B., Drescher, C., Hanuschkin, A., Winkler, L., Peters, S., Muller, K-L. Towards CRISP-ML(Q): A machine learning process model with quality assurance methodology. *Machine Learning Knowledge Extraction*, **3**(2), 392-413, 2021. DOI: 10.3390/make3020020.
31. Liu, S., Borovykh, A., Grzelak, L. A., Oosterlee, C. W. A neural network-based framework for financial model calibration. *Journal of Mathematics in Industry*, **9**, Art Id: 9, 2019. DOI: 10.1186/s13362-019-0066-7.
32. Bussmann, N., Giudici, P., Marnelli, D., Papenbrock, J. Explainable AI in fintech risk management. *Frontiers in Artificial Intelligence*, 3, Art Id:26, 2019. DOI: 10.3389/frai.2020.00026.